\documentclass[numreferences]{kluwer}
\usepackage{klucite}
\usepackage{epsfig}

\newcommand{\refb}[1]{(\ref{#1})}
\newcommand{\reftwo}[2]{(\ref{#1}-\ref{#2})}
\newcommand{\funcg}[2]{\mu_{#2,0} \, \mu_{0,#1-#2} \, a_{#2} \, a_{#1-#2}}
\newcommand{\funch}[2]{\sum_{j=0}^{#1-#2-1} {#1 - #2 \choose j} (-1)^j \, 
\mu_{#2+j}}

\begin{document}                                                            
\begin{article}

%%%
%%% Opening
%%%

\begin{opening}         

% Title block
\title{Exact Mean-Field Solutions of the\\Asymmetric Random Average Process} 
% Author block
\author{Frank \surname{Zielen}\email{fz@thp.uni-koeln.de}}
\author{Andreas \surname{Schadschneider}\email{as@thp.uni-koeln.de}}
\institute{Institut f{\"u}r Theoretische Physik, Universit{\"a}t zu K{\"o}ln,
50937 K{\"o}ln, Germany}
% Extra information
\runningtitle{Exact Mean-Field Solutions of the Asymmetric Random Average 
Process}
\runningauthor{Zielen and Schadschneider}
% Abstracts and Keywords
\begin{abstract}
We consider the asymmetric random average process (ARAP) with continuous 
mass variables and parallel discrete time dynamics studied recently by 
Krug/Garcia and Rajesh/Majumdar [both {\it Jrl.\ Stat.\ Phys.}~{\bf 99} 
(2000)]. The model is defined by an arbitrary state-independent fraction 
density function $\phi(r)$ with support on the unit interval. We examine 
the exactness of mean-field steady-state mass distributions in
dependence of $\phi$ and identify 
as a conjecture based on high order calculations
the class $\mathcal{M}$
of density functions yielding
product measure solutions.
Additionally the exact form of the 
associated mass distributions $P(m)$ is 
%% calculated.
derived.
Using these results we show examplary the exactness of the mean-field 
ansatz for monomial fraction densities $\phi(r)=(n-1) r^{n-2}$ with 
$n \geq 2$. For verification we 
%%derive
calculate
the mass distributions $P(m)$ 
explicitly and prove directly that product measure holds.
Furthermore we show that even in cases where the steady state is
not given by a product measure very accurate
%%approximations can be obtained.
approximants can be found in the class $\mathcal{M}$.
\end{abstract}

\end{opening}           

%%%
%%% Introduction
%%%

\section{Introduction}  

Interacting particle systems far from equilibrium represent due to their
wide applications in physics and other related topics like traffic
flow modeling \cite{schad:review} or force propagation in granular
media \cite{liu:forfbp} a popular theoretical research field 
\cite{Schuetz:review}. In this paper we focus on a model studied recently 
by Krug and Garcia \cite{KrugJ:asyps} and Rajesh and Majumdar
\cite{RajeshR:conmm}
%%. 
that is closely related to the $q$ model introduced by Coppersmith
{\it et al.} \cite{CoppersmithSN:forfbp}.
It describes the movement of particles on the
real line according to a given probability distribution depending on
the distance of two particles. This system is equivalent to a stick
model where the height of the sticks represents the particle gap
\cite{KrugJ:asyps,RajeshR:conmm}.
Below we present a brief description for completeness and demarcation
of our framework.

The asymmetric random average process (ARAP) is defined on a one-dimensional
periodic lattice with $L$ sites. Each site $i$ carries a
non-negative continuous mass variable $m_i$. In every time step $t\to t+1$
for each site a random number $r_i$ is generated from a time- and 
site-independent probability distribution $\phi$ defined on $[0,1]$. 
The fraction $r_i$ determines the amount of mass $r_i m_i$ transported 
from site $i$ to site $i+1$ (asymmetric shift). 

In the following we concentrate on the infinite system in the
thermodynamic limit, i.e. $L \rightarrow \infty$ and $M=\sum_i m_i
\rightarrow \infty$ with finite constant density 
$\rho=\frac{M}{L}$. 
Furthermore we examine only time independent steady state dynamics.

In case of a uniform fraction density $\phi(r)=1$ mass distributions
and moments have been calculated in mean-field approximation for
different types of dynamics 
\cite{KrugJ:asyps,majumdar:diffaggrfrag,RajeshR:conmm}. 
Although these
analytical results show excellent agreement with numerical
simulations, one can prove that for random sequential update a product
measure ansatz fails \cite{RajeshR:conmm}. For the fully parallel
update, however, one can adopt the result of \cite{CoppersmithSN:forfbp}
where the exactness of the mean-field approach is proven in the context of force
fluctuations in bead packs
described by the $q$ model
\cite{CoppersmithSN:forfbp,claudin:stressfluc}.
%% - a class of models closely related to the
%%parallel ARAP 

In this paper we like to reinforce investigations by considering
arbitrary state independent fraction density functions $\phi=\phi(r)$
and focussing on the fully parallel update only.
Such density functions are relevant for practical applications, e.g.\
as approximation for the ARAP with a finite mass cut-off 
\cite{ZieScha:cutoff}
%%.
or for describing suitable bead pack problems.

In section \ref{chap_crit1} a functional equation acting on the
Laplace-space of the single-site mass distributions $P(m)$ is derived.
Product measure holds if a mean-field ansatz is a solution of this
condition.

In section \ref{chap_crit2} we determine the set $\mathcal{M}$
of all fraction
probability densities $\phi(r)$ yielding product measure states.
This result represents a conjecture based on exact high order calculations.
 Now statements about the exactness of mean-field are possible without
solving the master equation. We also derive the corresponding
mass distributions $P(m)$.

As an example the ARAP with monomial $\phi$-function is studied
%%.
(section \ref{chap_example}).
We calculate $P(m)$ and prove the exactness of mean-field by using the
explicit form of $P(m)$ on the one hand and the criterion of the last
section on the other hand.
Due to the fact that the $n$ dimensional $q$ model with uniform distributed $q$'s leads to the same mass distribution \cite{CoppersmithSN:forfbp} its relationship with the ARAP is discussed, too.

In section \ref{chap_applic} we demonstrate that the results obtained
in the previous sections can be used to obtain very accurate approximations
for the mass distributions of ARAPs which do not lead to product
measures.

In section \ref{chap_ende} we conclude with a
%%brief discussion.
summary and give a brief outlook for further work.
%%

%%%%%%%%%%%%%%%%%%%%%%%%%%%%%%%%%%%%%%%%%%%%%%%%%%%%%%%%%%%%%%%%%%%%%%%%%%%%%
%%%
%%% A Functional Equation of Exact Mean-Field Solutions
%%%

\section{A Functional Equation of Exact Mean-Field Solutions} 
\label{chap_crit1}

In this section we derive a functional equation (see eq.~\refb{cond2_Q}
below) for determining and testing mean-field solutions.

The fundamental element of all upcoming considerations is the master equation.
For the single site mass distribution $P(m)$ in the stationary
state it reads \cite{RajeshR:conmm} 
\begin{eqnarray} \label{mgl_singleP}
P(m_2^\prime) & = & \int_0^\infty \!\! dm_1 \int_0^\infty \!\! dm_2 \; 
P(m_1,m_2) \int_0^1 dr_1 \int_0^1 dr_2 \; \phi(r_1) \phi(r_2) \nonumber \\  
& & \times \; \delta \left( m_2^\prime - \left[ r_1 m_1 + (1-r_2) m_2 \right] 
\right) \;.
\end{eqnarray}
The $\delta$-function ensures mass conservation. We have assumed 
translational invariance so that the distribution is site-independent.

By the mean-field
ansatz $P(m_1, m_2, \ldots) = \prod_i P(m_i)$ equation
\refb{mgl_singleP} determines the single site distribution $P(m)$. The
resulting product measure is exact if the mean-field ansatz holds for
all joint probabilities in the stationary state, too, i.e. $P(m)$ has 
to satisfy
\begin{eqnarray} \label{mgl_allP}
\prod_{i=2}^k P(m_i^\prime) & = & \left\{\prod_{i=1}^k \int_0^\infty \!\! 
dm_i \; P(m_i) \int_0^1 \!\! dr_i \; \phi(r_i) \right\} \nonumber \\
& & \times \prod_{i=2}^k \delta \left( m_i^\prime - \left[ r_{i-1} m_{i-1} 
+ (1-r_i) m_i \right] \right) 
\end{eqnarray}   
for all $k \in \dN_{\geq 2}$. For $k=2$ this equation reduces 
to \refb{mgl_singleP}. This appears to be an infinite set of conditions, 
but Laplace transforming $P(m_1, m_2, \ldots)$ reduces \refb{mgl_allP} 
to just one functional equation.
By introducing the $k-$dimensional Laplace-transform
\begin{equation}
Q(s_1,\ldots,s_k) \equiv \int_0^\infty \!\! d^km \; P(m) e^{-(m,s)},
\end{equation}
where $(m,s)=\sum_{i=1}^k m_i s_i$, and using the map
\begin{equation}
F_Q(s,\tilde{s}) \equiv \int_0^1 \!\! dr \; \phi(r) Q\left( \left(1-r\right) 
s + r \tilde{s} \right)
\end{equation} 
equation \refb{mgl_allP} reads in Laplace space
\begin{equation} \label{cond_allQ}
\prod_{i=1}^k Q\left(s_i\right) = F_Q(0,s_1) \cdot \prod_{i=1}^{k-1} 
F_Q(s_i,s_{i+1}) \cdot F_Q(s_k,0) 
\end{equation}
for $k \in \dN$.

By a straightforward proof we now show that the conditions for $k \neq 2$ 
are redundant. The $k=1$-equation
\begin{equation} \label{cond1_Q}
Q(s) = F_Q(s,0) \cdot F_Q(0,s)
\end{equation}
is used to determine $Q$. We rewrite the $k=2$-criterion 
using (\ref{cond1_Q}) and obtain
\begin{equation} \label{cond2_Q}
F_Q(s_1,s_2) = F_Q(s_1,0) \cdot F_Q(0,s_2) \;.
\end{equation}
Applying \refb{cond1_Q} and \refb{cond2_Q} proves the validity of 
\refb{cond_allQ} for all $k \geq 3$ and using the identity 
$Q(s)=F_Q(s,s)$ equation \refb{cond1_Q} becomes a special case of 
\refb{cond2_Q}. So \refb{cond2_Q} is the only necessary equation to
determine a mean-field solution $(s_1=s_2)$ and check its accuracy 
$(s_1\neq s_2)$. 

%%%%%%%%%%%%%%%%%%%%%%%%%%%%%%%%%%%%%%%%%%%%%%%%%%%%%%%%%%%%%%%%%%%%%%%%%%%%%
%%%
%%% Exact Mean-Field Solutions
%%%

\section{Exact Mean-Field Solutions} \label{chap_crit2}

In this section we derive the set of density functions $\phi(r)$ that
result in product measure steady states. This yields a more useful
criterion for determining the exactness of a mean-field solution
without calculating and verifying $Q(s)$ by condition \refb{cond2_Q}.
In addition we derive the mass distribution $P(m)$.

We start by proving that equation \refb{cond1_Q} has
always a unique solution in the space of functions that are
analytical in the origin. Intuitively one supposes this feature
because the mass moments 
\begin{equation}
m_n \equiv \left\langle m^n \right\rangle_{P(m)}=\int_0^\infty dm\, m^n P(m)
\label{def_moments}
\end{equation}
are (formally) generated by $m_n = (-1)^n Q^{(n)}(0)$. But a priori we
do not know if all derivatives $Q^{(n)}$ of the moment function $Q(s)$ 
exist or ensure convergence. The moments $m_0=1$ and $m_1=\rho$ are
determined by the normalization and the density $\rho$, respectively.

We first represent the moment function as a (formal) 
power series, i.e. $Q(s) = \sum_{n} a_n s^n$. The coefficients
$a_n$ are related to the moments (\ref{def_moments}) by
$a_n=(-1)^n\frac{m_n}{n!}$ and thus we have
$a_0=1$ and $a_1=-\rho$. The
remaining coefficients are determined with help of a recurrence
relation obtained by inserting the series representation into
\refb{cond1_Q}:
\begin{equation} \label{cond1_series}
a_n = \frac{1}{1-\mu_{n,0}-\mu_{0,n}}\sum_{k=1}^{n-1} \mu_{k,0} \, 
\mu_{0,n-k} \, a_k \, a_{n-k} \quad\qquad (\forall{n \geq 2}).  
\end{equation}
Here $\mu_{n,m}$ are generalized moments of the fraction density $\phi$ 
defined by 
\begin{equation} \label{def_mu}
\mu_{n,m} \equiv \left\langle r^n (1-r)^m \right\rangle_{\phi(r)} = 
\int_0^1 dr \phi(r) r^n (1-r)^m \;.
\end{equation}
We assume $\mu_{n,m}>0$ in the following which is equivalent to
$\phi(r) \neq 0$ for $r\in(0,1)$. $\mu_{n,m}=0$ does not occur for
continuous distributions, but e.g.\ for 
$\phi(r)=p \delta(r) + (1-p) \delta(1-r)$. 
This ARAP is trivial for $p=0$  or $p=1$ and not solvable under mean-field 
assumptions for $0<p<1$. From
\begin{equation} \label{relation_mu}
1-\mu_{n,0}-\mu_{0,n}\ \stackrel{\refb{def_mu}}{=} \ 
\sum_{k=1}^{n-1} {n \choose k} \mu_{k,n-k} > 0 
\quad \qquad (\forall n \geq 2)
\end{equation} 
we conclude that all $a_n$ are well-defined and the solution of
\refb{cond1_Q} is unique. By the formula of Cauchy-Hadamard we then show
that $Q$ is holomorphic in $s=0$:
We start by proving the lemma
\begin{equation} \label{ana_lemma2}
\left| a_n \right| \leq D^{n-1} c_n \rho^n 
\quad\qquad (\forall n \geq 1)
\end{equation}
with $D \equiv \frac{1}{1-\mu_{2,0}-\mu_{0,2}}$ and the density $\rho$. 
Here $\left\{c_n\right\}_{n \in \dN}$ are the Catalan numbers 
\cite{WilfH:generating} fulfilling the equations
\begin{equation} \label{catalan_numbers}
c_1 = 1 \quad {\rm and}\quad 
c_n = \sum_{k=1}^{n-1} c_k c_{n-k} = \frac{1}{n}{2(n-1) 
\choose n-1} \quad (\forall n \geq 2).
\end{equation}
Inserting \refb{ana_lemma2} into \refb{cond1_series} 
using \refb{catalan_numbers} and the fact that $\mu_{n,m} > \mu_{n+k_1,m+k_2}$ 
for all $k_i \in \dN$ shows inductively the validity of the lemma 
(\ref{ana_lemma2}). 
From $\lim_{n\to\infty} \sqrt[n]{c_n} = 4$ we conclude that the
series expansion $Q(s)=\sum_n a_ns^n$ has a positive radius of convergence. 

After proving that mean-field solutions are always representable
as power series we now try to express the exact solutions in terms of
the density function $\phi$. Inserting $Q(s)=\sum_n a_n s^n$ into
\refb{cond2_Q} yields an infinite set of conditions
\begin{equation} \label{cond2_series}
{n \choose k} \mu_{k,n-k}  \, a_n = \mu_{k,0} \, \mu_{0,n-k} \, a_k \, 
a_{n-k} 
\quad\ \ (\forall n \in \dN_0, \;\; \forall k \in \left\{0,\ldots,
n\right\}).
\end{equation}
In the following we refer to a special equation of \refb{cond2_series} 
as $(n,k)$-condition or equation of order $n$.  
Summing over $k=1,\ldots,n-1$ in \refb{cond2_series} yields 
condition \refb{cond1_series}. This shows again that \refb{cond2_Q} includes
formula \refb{cond1_Q} and implies convergence of $Q(s)$.

Now we are confronted with the problem that $a_n$ is determined by
$n+1$ equations. Since the $(n,0)$- and $(n,n)$-conditions match
identities there are effectively $n-1$ equations to be fulfilled for
$n\geq2$. Thus for $n\geq3$ the occurence of inconsistencies is
possible and for arbitrary $\phi$ or moments $\mu_{n,m}$ we see by
explicit calculation contradictions in order $n=3$ already. Is it
possible to find a set of moments $\left\{ \mu_{n,m} \right\}$ such that
\refb{cond2_series} is satisfied for all $(n,k)$?

Assuming that all $(n,k)$-conditions yield the same $a_n$ we see
from (\ref{cond2_series}) that the function
\begin{equation} \label{cond_mu}
f(n,k) \equiv 
\frac{\mu_{k,0} \, \mu_{0,n-k} \, a_k \, a_{n-k}}{{n \choose k} \mu_{k,n-k}}
\;.
\end{equation}
is independent of $k$. Therefore the $n-2$ consistency equations 
\begin{equation} \label{def_f_n}
f(n,1)=f(n,2)=\ldots =f(n,n-1)\equiv a_n
\end{equation}
have to be satisfied.
The definition (\ref{def_mu}) implies that
$\mu_{n,m}$ can be expressed by $\mu_{j,0}$ only:
\begin{equation} \label{def_mu_n}
\mu_{n,m}=\sum_{j=0}^m{m \choose j}(-1)^j \mu_{n+j}\quad\quad{\rm with} 
\quad\quad \mu_j\equiv \mu_{j,0}.
\end{equation}
Together with (\ref{cond2_series}) it follows inductively 
that $a_n=a_n(\mu_0,\ldots,\mu_n)$. This leads to a
successive constructive approach in $n$: We try to find a $\mu_n$ (depending
on $\mu_0,\ldots,\mu_{n-1}$) that solves all conditions (\ref{def_f_n})
of order $n$ starting
with $n=3$ and repeat this for all upcoming orders $n=4,5,\ldots\;$.

The $n-2$ equations \refb{def_f_n} are linear in $\mu_n$.
This ensures uniqueness 
of a possible solution. For arbitrary $k,\tilde{k}=1,\ldots, n-1$
with $k\neq \tilde{k}$ we obtain
\begin{equation} \label{cond_mu_expl}
\mu_n = \frac{
{n \choose k} \, h_{n,k} \, g_{n,{\tilde{k}}} - {n \choose \tilde{k}}
  \, h_{n,\tilde{k}} \, g_{n,k}}{ {n \choose \tilde{k}} (-1)^{n-\tilde{k}}
  \, g_{n,k} - {n\choose k} (-1)^{n-k} \, g_{n,{\tilde{k}}}}\;.
\end{equation}
%%The functions
Here
$g_{n,k} \equiv \funcg{n}{k}$ and 
$h_{n,k} \equiv \funch{n}{k}$ only depend on $\mu_1,\ldots,\mu_{n-1}$.
Thus (\ref{cond_mu_expl}) gives us the desired recursion relation to
determine all moments $\mu_n$. However, we have to check whether the
r.h.s.\ of \refb{cond_mu_expl} is independent of $k$ and $\tilde{k}$.
We 
%%checked 
have done
this up to order $n=10$ by computer algebra and conjecture 
the validity for all $n$. Our results furthermore lead us to
conjecture the following form of the solution of the density moments:
\begin{equation} \label{moment_criterion}
\mu_n = \prod_{l=0}^{n-1} \frac{l + \lambda_1}{l + \lambda_2} 
= \frac{\Gamma(n+\lambda_1)}{\Gamma(\lambda_1)}
\frac{\Gamma(\lambda_2)}{\Gamma(n+\lambda_2)} 
\end{equation}
with
\begin{equation} 
\label{moment_criterion_lambda}
\lambda_1 = \mu_1 \frac{\mu_1 - \mu_2}{\mu_2 - \mu_1^2} \;, \qquad
\lambda_2 = \frac{\mu_1 - \mu_2}{\mu_2 - \mu_1^2} \;, 
\end{equation}
which again has been checked up to $n=10$.
Note that $\mu_1$ and $\mu_2$ are free parameters that only have to 
be choosen with respect to the general moment properties
\begin{equation} \label{parameter_space}
1>\mu_1>\mu_2\geq\mu_1^2 \;.
\end{equation}
The special case $\mu_1=\mu_2^2$ yields $\mu_n=\mu_1^n$ representing
$\phi(r)=\delta(r-\mu_1)$ which leads to $Q(s)=\exp(-\rho s)$ and thus
to the mass distribution $P(m)=\delta(m-\rho)$.
So we assume $\mu_2 > \mu_1^2$ in the following. Under these
restrictions $\mu_{n \geq 3}$ is pole-free and satisfies
$0<\mu_{n+1}<\mu_n$ as demanded.

(\ref{moment_criterion})-(\ref{parameter_space}) define the set 
$\mathcal{M}$ of all fraction densities $\phi(r)$ yielding product 
measure steady-state distributions. So equations
\reftwo{moment_criterion}{moment_criterion_lambda} represent a
powerful criterion for determining the exactness of a mean-field
ansatz: We only have to calculate the moments $\mu_n$ of the fraction
density (if they are not already given) and check consistency with
(\ref{moment_criterion}) and (\ref{moment_criterion_lambda}) - without even
calculating the mean-field mass distribution or its Laplace-transform!

Note that the parametrization of $\mathcal{M}$ in terms of $\mu_1$ and
$\mu_2$ is arbitrary and a consequence of our construction - other
parametrizations are possible. For symmetric densities
$\phi(r)=\phi(1-r)$ the space $\mathcal{M}$ reduces to one dimension because
$\mu_1=\frac{1}{2}$ is fixed.

After determining the class $\mathcal{M}$ of fraction densities
leading to a product measure we now like to calculate the single 
site mass distribution $P(m)$ for these 
$\phi \in \mathcal{M}$. From the $(n+1,n)$-condition
\refb{cond2_series} we derive the recurrence relation
\begin{equation} \label{mf_recurrence}
\lambda_2 \, (n+1) a_{n+1} + \rho \, (n + \lambda_2) a_n = 0 
\end{equation}
that is valid for all $n \in \dN_0$. 
\refb{mf_recurrence} corresponds to a first order differential equation for 
the moment function $Q(s)=\sum_n a_n s^n$:
\begin{equation}
\left( \lambda_2 + \rho s \right) Q^\prime(s) + \lambda_2 \rho \, Q(s) = 0 \;.
\end{equation}
Using the boundary condition $Q(0)=1$ we obtain
\begin{equation} \label{mfsol_Q}
Q(s) =\frac{1}{\left(1 +\frac{\rho}{\lambda_2} s\right)^{\lambda_2}}
\end{equation}
or, by calculating the inverse Laplace-transform,
\begin{equation} \label{mfsol_P}
P(m) = \frac{\lambda_2^{\lambda_2}}{\Gamma\left( \lambda_2 \right)} 
\frac{1}{\rho} \left( \frac{m}{\rho} \right)^{\lambda_2-1} 
e^{-\lambda_2 \frac{m}{\rho}} \;.
\end{equation}
In contrast to $\mu_n$ depending on both $\lambda_1$ and $\lambda_2$
-- see equation \refb{moment_criterion} -- $P(m)$ is a function of 
$\lambda_2$ only.
So ARAPs with $\lambda_2$ fixed and $\lambda_1$ arbitrary have
identical mass distributions \refb{mfsol_P}.
  
%%%%%%%%%%%%%%%%%%%%%%%%%%%%%%%%%%%%%%%%%%%%%%%%%%%%%%%%%%%%%%%%%%%%%%%%%%%%%
%%%
%%% Explicit solution for monomial density function $\phi(r)$
%%%

\section{Explicit solution for monomial density function $\phi(r)$} 
\label{chap_example}

In this section we derive the solution of the ARAP with density function
\begin{equation} \label{densityfunction}
\phi_n(r)=(n-1) r^{n-2} \quad\qquad (\forall n \in \dN_{\geq2})
\end{equation}
in a closed form ($n-1$ is the normalization
constant) and prove the exactness of
the product measure $\prod_k P(m_k)$ both explicitly and using
the criterion
%%(\ref{moment_criterion}), (\ref{moment_criterion_lambda}).
(\ref{moment_criterion}-\ref{moment_criterion_lambda}).
Additionally a brief comparison between ARAP and the $q$ model is presented.

We start by constructing the analytic solution of the functional equation
\refb{cond1_Q} explicitly. For $n=2$, where $\phi_n$ reduces to a
constant distribution, we refer to
\cite{CoppersmithSN:forfbp,KrugJ:asyps,RajeshR:conmm} and find
\begin{equation} \label{n2_solution}
Q_2(s)=\frac{1}{\left(1+\frac{\rho}{2} s \right)^2} \;.
\end{equation}
To solve the $n=3$-problem we generalize the method used in 
\cite{CoppersmithSN:forfbp,KrugJ:asyps,RajeshR:conmm}. 
Defining the functions
\begin{equation} \label{def_VW}
V(s) = \int_0^1 dr \, Q_3(rs) \quad \mbox{and} \quad
W(s) = \int_0^1 dr \, r \, Q_3(rs)
\end{equation}
the functional equation \refb{cond1_Q} transforms into
\begin{equation} \label{cond1_VW}
Q_3(s) = 4 W(s) \left( V(s) - W(s) \right) \;.
\end{equation}
From \refb{def_VW} we derive
\begin{equation} \label{equQ}
s V'(s) + V(s) = Q_3(s) = s W'(s) + 2 W(s)
\end{equation}
which implies the following relation between $V$ and $W$:
\begin{equation} \label{equV}
V(s) = W(s) + \frac{1}{s} \int W(s) ds \;.
\end{equation}
Defining the antiderivative $f(s)\equiv\int W(s) ds$ and inserting 
\refb{equV} and \refb{equQ} into \refb{cond1_VW} yields the nonlinear 
differential equation \begin{equation} \label{cond1_dgl}
s^2 f''(s) + 2 s f'(s) - 4 f'(s) f(s) = 0
\end{equation}
with boundary conditions $f'(0)=\frac{1}{2}$ and
$f''(0)=-\frac{1}{3}\rho$. This results in $f(s)=\frac{s}{2} \left( 1+
  \frac{\rho}{3} s\right)^{-1}$, e.g. solved by power series ansatz, and we get
\begin{equation} \label{n3_solution}
Q_3(s) = \frac{1}{\left( 1 + \frac{\rho}{3} s \right)^3} \;.
\end{equation}
The results \refb{n2_solution} and \refb{n3_solution} suggest
the assumption
\begin{equation} \label{n_solution}
Q_n(s) = \frac{1}{\left( 1 + \frac{\rho}{n} s \right)^n} \; ,
\end{equation}
for general $n$.
$Q_n$ fulfills the initial conditions $Q_n(0)=1$ and $Q_n'(0)=-\rho$. By a 
straightforward induction in $n$ (using partial integration) we are able 
to prove
\begin{equation}\label{help}
F_{Q_n}(s,0) = \frac{1}{1 + \frac{\rho}{n} s } \quad {\rm and} \quad
F_{Q_n}(0,s)  = \frac{1}{\left(1 + \frac{\rho}{n} s\right)^{n-1}}
\end{equation}
and see with \refb{n_solution} that \refb{cond1_Q} is valid.

%%
%%We like to mention that $Q_n$ in addition to the condition
%%\refb{cond1_Q} also satisfies the functional equation 
%%$Q(s) = (F_Q(s,0))^n$ considered in \cite{CoppersmithSN:forfbp}.
%%

The next step is to verify the functional equation \refb{cond2_Q}.
This is done again straightforwardly by induction in $n$. So
\refb{n_solution} represents the exact solution of the ARAP with
fraction density \refb{densityfunction}.

%%We now
Now we
rederive (\ref{n_solution}) using the mean-field criterion
\reftwo{moment_criterion}{moment_criterion_lambda}. Calculating the 
moments of $\phi_n$ exactly is an easy task and yields
\begin{equation} \label{moments_phin}
\mu_k = \frac{n-1}{n-1+k} \;.
\end{equation}
Especially we have $\mu_1 = \frac{n-1}{n}$ and $\mu_2 =
\frac{n-1}{n+1}$ and, using (\ref{moment_criterion_lambda}), we obtain
$\lambda_1=n-1$ and $\lambda_2=n$. 

The exact form (\ref{moments_phin})
of the moments is reproduced by taking into account
(\ref{moment_criterion}).
Thus the monomial density functions $\phi_n$ are elements of
$\mathcal{M}$. 
%% In addition, $\lambda_2=n$ shows the equivalence of \refb{mfsol_Q} and \refb{n_solution}.
Using $\lambda_2=n$ shows the equivalence of \refb{mfsol_Q} and \refb{n_solution} after all.

Additionally this example confirms the validity of the conjecture (\ref{moment_criterion}-\ref{moment_criterion_lambda}).

For completeness we also give the explicit form of the single-site mass
distribution for the monomial density functions \refb{densityfunction}:
\begin{equation} \label{n_mass_distribution}
P_n(m) = \frac{n^n}{(n-1)!} \frac{m^{n-1}}{\rho^n} e^{-n \frac{m}{\rho}} \;.
\end{equation}

Finally we like to mention that $Q_n$ also satisfies the relation
\begin{equation} \label{copper_equation}
Q_n(s) = (F_{Q_n}(0,s))^n = \left[ \int_0^1 dr \phi(r) Q_n(rs) \right]^n \;.
\end{equation}
This functional equation represents the master equation of the $n$ ancestor $q$ model with uniform distributed $q$'s and was explicitly solved in \cite{CoppersmithSN:forfbp}. Because of the formal difference of the underlying equations (\ref{cond1_Q}) and (\ref{copper_equation}) the coincidence of the corresponding solution (\ref{n_solution}) is remarkable. Otherwise this result is pointing to a deeper relationship between $n$ dimensional $q$ model and ARAP.

As already noticed in \cite{KrugJ:asyps} the two ancestors $q$ model ($n=2$) corresponds to an ARAP with symmetric density function $\phi(r)=\phi(1-r)$. Coppersmith {\it et al.} \cite{CoppersmithSN:forfbp} identified a set of mean-field solutions generated by monomial distributions, i.e. $f(q_{ij})=q^m$, yielding $\phi(r)=\frac{(2m+1)!}{(m!)^2}r^m(1-r)^m$ and it is easy to see that these densities are also elements of $\mathcal{M}$.
%%

%%%%%%%%%%%%%%%%%%%%%%%%%%%%%%%%%%%%%%%%%%%%%%%%%%%%%%%%%%%%%%%%%%%%%%%%%%%%%
%%%
%%% Approximative mass distributions for arbitrary density functions
%%%

\section{Approximative mass distributions for arbitrary density functions}
\label{chap_applic}

As an application of our results, we construct approximative mass 
distributions for arbitrary density functions $\phi$. 
This is done by calculating the
parameter $\lambda_2=\lambda_2(\phi)$ with the help of
\refb{moment_criterion_lambda} (using the exact moments $\mu_1$ and
$\mu_2$ of $\phi$) and taking the corresponding mean-field
solution \refb{mfsol_P} as an approximation. To illustrate this method
and estimate its quality we consider two examples.

Our discussion starts with density functions being convex combinations of 
elements of $\mathcal{M}$. We restrict here to the special case
\begin{equation} \label{convex_density}
\phi_c = (1-c) \phi_2 + c \phi_3\;, \quad c \in [0,1]
\end{equation}
with the monomial density functions $\phi_n$ defined in 
\refb{densityfunction}. This convex combination of probability densities 
conserves their basic properties like normalization or positivity.  
Calculating the first and second
moment of $\phi_c$ yields  $\lambda_2 = \frac{6}{3-c^2}$.
Inserting this result into \refb{mfsol_P} generates $c$-dependent
approximations $P_c$.

\begin{figure}[h]
\centerline
{
\epsfig{file=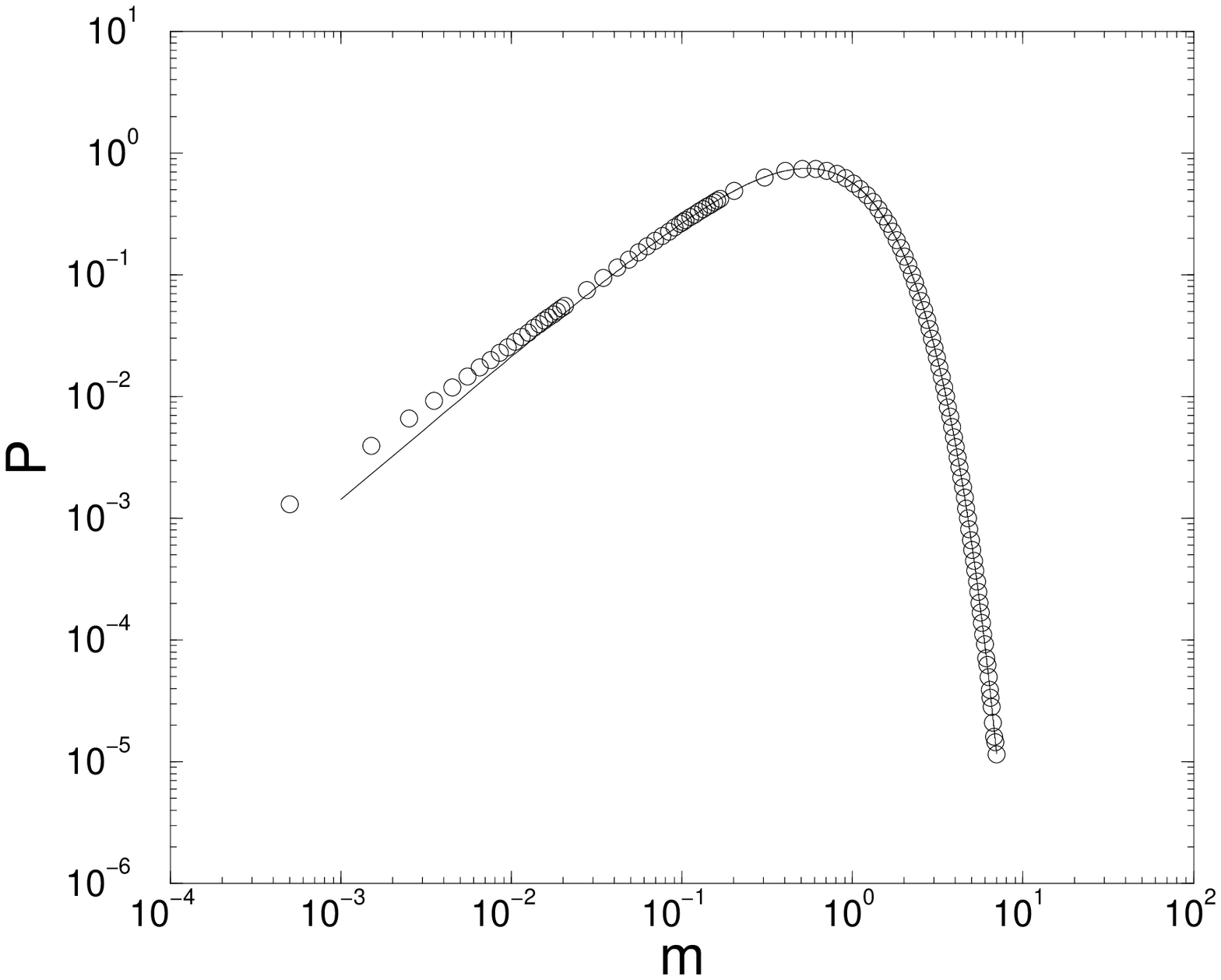, height=7cm}
\epsfig{file=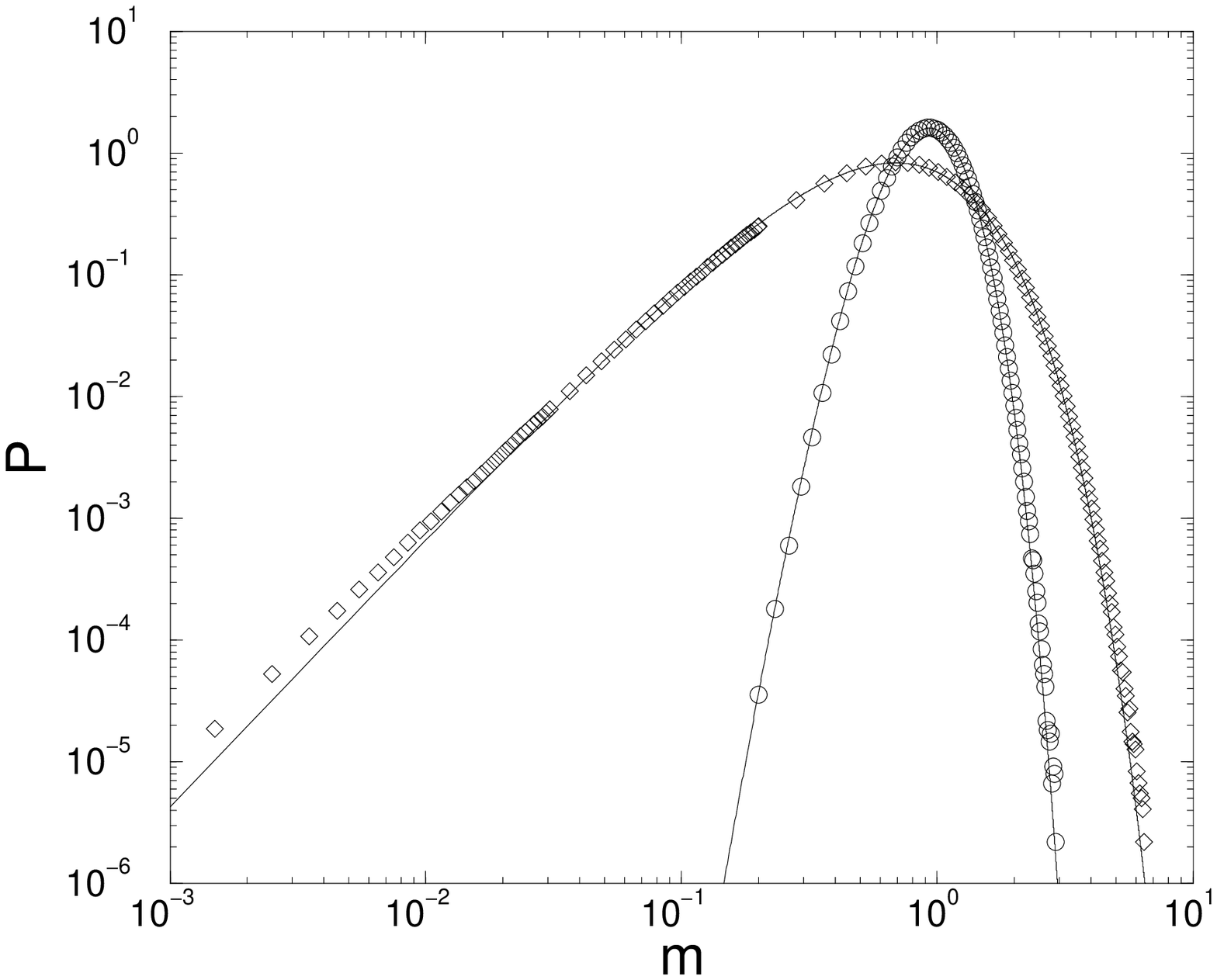, height=7cm}
}
\vspace{1cm}
\caption{} \label{approx_fig}
\end{figure}

Comparing the distribution $P_c$ with numerical data shows
an excellent agreement between approximation and the results of
Monte Carlo simulations for all values of $c$ (see figure \ref{approx_fig}).
Only for small masses $m$ systematical differences occur.

Furthermore one can prove that $\phi_c \not\in \mathcal{M}$ for all
$0<c<1$. This is most easily seen by comparing the third moment $\mu_3$ of
\refb{convex_density} with the corresponding mean-field expression
\refb{moment_criterion}. So the excellent agreement of the
data match is far from trivial.
Nevertheless $\phi_c$ is an interpolation between exact mean-field
solutions and may inherit some of their properties.

Our second example represents the simplest version of an ARAP with
cut-off. It is based on the model with uniform fraction density, but
enhanced by an additional parameter $r_0 \in [0,1]$. The cut-off $r_0$
controls the movement in the following way: A stick fragment $r_i m_i$
is only transferred if $r_i \leq r_0$. The corresponding density
function takes the form:
\begin{equation} \label{cutoff_density}
\phi_{r_0}(r) = (1-r_0) \delta(r) + H(r_0-r) 
\end{equation} 
where $H(r)$ is the Heaviside step-function.
A detailed discussion of several truncated ARAPs can be found in 
\cite{ZieScha:cutoff}.

If $r_0=1$ we obtain the free ARAP which is exactly solvable by product 
measure ansatz \cite{CoppersmithSN:forfbp}. For $0<r_0<1$ one can show 
that $\phi_{r_0} \not\in \mathcal{M}$. In the case $r_0=0$, where no
motion is possible, the mean-field condition \refb{cond2_Q} reduces to 
an identity - so any distribution represents
an exact mean-field solution. In this sense \refb{cutoff_density}
interpolates between ARAPs with product measure steady state as in the
first example, but the construction is not a convex combination.

The approximants $P_{r_0}$ are calculated as described above and match
the Monte-Carlo simulation data perfectly again
except deviations for small $m$
(see figure \ref{approx_fig}).

%%%%%%%%%%%%%%%%%%%%%%%%%%%%%%%%%%%%%%%%%%%%%%%%%%%%%%%%%%%%%%%%%%%%%%%%%%%%%
%%%
%%% Conclusions
%%%

\section{Conclusions} \label{chap_ende}

We have studied the asymmetric random average process with arbitrary
fraction density $\phi$. Based on the analysis of the functional
equation (\ref{cond2_Q}) we have identified all ARAPs where the
stationary mass distribution $P(m)$ is given by a product measure. The
corresponding $\phi$-functions form the mean-field class
$\mathcal{M}$. They are given by their moments $\mu_n=\langle
r^n\rangle_\phi$ which depend on two free parameters, $\lambda_1$ and
$\lambda_2$, related to the first moments $\mu_1$ and $\mu_2$
(\ref{moment_criterion_lambda}). These can be chosen arbitrarily
subject only to the general conditions (\ref{parameter_space}). For
the product measure ARAPs also the stationary mass distribution $P(m)$
can be calculated explicitly (\ref{mfsol_P}).  Surprisingly it depends
only on the parameter $\lambda_2$.

The presented approach is rigorous except for a formal proof of
(\ref{moment_criterion}) which we have conjectured on the basis of
computer algebra calculations.

As shown our results can be used to obtain accurate approximations for
ARAPs which do not lead to exact product measures: instead of
calculating mean-field approximations by \refb{cond1_Q} one can use
the corresponding distribution \refb{mfsol_P} with appropriately
determined $\lambda_2=\lambda_2(\phi)$ as a first guess. A detailed
discussion of $\mathcal{M}$ could explain why this method often matches
perfectly with numerical simulation data. We suppose that
arbitrary choosen $\phi$ lie "near" to the subspace $\mathcal{M}$ with
respect to a suitable chosen norm.

We like to mention that our approach can easily be adopted to the ARAP
with random sequential update. Since a mean-field ansatz breaks down
even for the simple model with uniform $\phi$-function
\cite{RajeshR:conmm} we 
%%have renounced investigations here.
%%
have not done calculations for this case.
Nevertheless the existence of product measure ARAPs with continuous
time dynamics is also possible.
Furthermore one could try to transfer our calculations to the $q$ model to supplement the findings of \cite{CoppersmithSN:forfbp}. It
%%
%%Furthermore it
would be interesting
to use our results for the calculation of other properties,
too,
e.g. the tracer diffusion coefficient \cite{SchuetzG:tradc}.

The calculation of exact solutions of ARAPs with arbitrary $\phi(r)$
seems to be the final step of work. Vanishing of the
2-site-correlation functions \cite{KrugJ:asyps} and the excellent
quality of mean-field approximations have already pointed to a
distribution that deviates only slightly from product measure
form. Supplementary our work gives by equation (\ref{mfsol_P}) a rough
shape of the general solutions.

Finally it would be interesting to extend our considerations to state
dependent density functions, i.e. $\phi=\phi(r,\{m_i\})$, and especially
to local probabilities $\phi(r,m)$.
These models are for example interesting for truncated asymmetric
random average processes \cite{ZieScha:cutoff}.

\begin{acknowledgements}
We would like to thank J. Krug and G. Sch\"utz for very useful discussions.
\end{acknowledgements}

%%%
%%% References
%%%

\bibliographystyle{klunum}
\bibliography{ZielenF}

%%%
%%% Figure Captions
%%%

\newpage
\pagestyle{empty}

{\large \bf Figure captions (Zielen, Schadschneider, JSP 101-110)}
\vspace{1cm}

{\it Figure 1.} Analytical $(-)$ and numerical $(\diamond,\circ$) mass 
distributions $P(m)$ of the convex combined ARAP ($c\!=\!0.5$, left diagram) and the truncated ARAP ($r_0\!=\!0.5$ $(\diamond)$ and $r_0\!=\!0.1$ $(\circ)$, right diagram).
The analytical curves are appropriate approximants 
taken from the mean-field class $\mathcal{M}$. The numerical results are 
obtained by Monte-Carlo simulations of systems with size $L=1000$
and random initial condition. After  
$10^4$ steps the distribution was measured for $10^7$ timesteps. A log-log plot is used to exhibit the deviations for small masses $m$ which can hardly be seen in a conventional representation.

%%%
%%% Figures
%%%

\newpage

{\large \bf Figure 1 (Zielen, Schadschneider, JSP 101-110)}
\vspace{1cm}

\end{article}

\end{document}